\documentclass{article}

\usepackage{graphicx}
\usepackage{wrapfig}
\usepackage{amsmath,amssymb,fontenc, times, mathptmx}
\usepackage[dvips]{color}
\usepackage{latexsym,graphicx,epsfig}

\def\lb{\linebreak[4]}
\newcommand{\be}{\begin{equation}}
\newcommand{\ee}{\end{equation}}
\newcommand{\bea}{\begin{eqnarray}}
\newcommand{\eea}{\end{eqnarray}}
\newcommand{\bes}{\begin{subequations}}
\newcommand{\ees}{\end{subequations}}
\newcommand{\bear}{\begin{equation}\begin{array}}
\newcommand{\eear}[1]{\end{array}\label{#1}\end{equation}}
\newcommand{\ls}[1]{\lambda_{#1}}

\newcommand{\fr}[2]{\dfrac{{ #1}}{{ #2}}}

\newcommand{\la}{\langle}
\newcommand{\ra}{\rangle}
\newcommand{\fn}[1]{\footnote{{#1}}}
\newcommand{\bu}{$\bullet$\ }
\renewcommand{\le}{\leqslant}
\renewcommand{\ge}{\geqslant}

\def\vep{{\varepsilon}}
\newcommand{\epe}{\mbox{$e^+e^-\,$}}

\def\cl{\centerline}

{\end{list}}
\newcounter{enumct}
\newenvironment{Enumerate}{\begin{list}{\arabic{enumct}.}%
{\usecounter{enumct}\setlength{\topsep}{0.2mm}%
\setlength{\partopsep}{0.2mm}\setlength{\itemsep}{0.2mm}%
\setlength{\parsep}{0.2mm}\setlength{\leftmargin}{4mm}}}
{\end{list}}

\newcommand{\missET}{\slash{\hspace{-2.4mm}E}_T}

\begin{document}
\title{Measuring mass and spin of Dark Matter particles with the aid energy spectra of single lepton  and dijet at the $e^+e^-$ Linear Collider
}

\author{I. F. Ginzburg\\
 Sobolev Institute of Mathematics,  Novosibirsk, 630090,  Russia;\\
Novosibirsk State University, Novosibirsk, 630090,  Russia\\
E-mail: ginzburg@math.nsc.ru}

\date{
}

\maketitle

\begin{abstract}
{In many models stability of  Dark Matter particles $D$  is
ensured by conservation of a new quantum number referred to as
$D$-parity. Our models  also contain  charged
$D$-odd particles $D^\pm$ with the same spin as $D$.

Here we propose a method to precisely measure the  masses
and spins of $D$-particles  in the process $\epe\to
D^+D^-\to DDW^+W^-\to DD (q\bar{q})(\ell\nu)$ with a signature {\it dijet
+ $\mu$ (or $e$) +  nothing}. It is shown that
 the energy distribution of the  lepton has
singular points  (upper edge and kinks or a peak) whose positions are
kinematically determined. For precise  determination  of $D$ and
$D^\pm$ masses, it is sufficient to measure these singular points and upper edge of dijet energy spectrum.
After this procedure, even an approximate measurement
of the corresponding cross section allows a determination of the spin
of $D$ particles to be performed.

 New points of this work are: 1)  We
propose to use only the well measurable energy
spectra of {\bf individual leptons} and the upper edge of the dijet energy spectrum. 2) We propose to
identify the spin of $D$-particles via
value of  the cross section  for the discussed
process.

}
\end{abstract}


\section{Introduction}\label{secintro}

\subsection{Models}\label{secmod}

In the   broad class of models Dark Matter (DM) consists of
particles $D$ similar to those in SM, with the following
properties:
\begin{Enumerate}
\item
The neutral DM particle $D$ with mass $M_D$ and spin $s_D=0$ or $1/2$
has a new conserved  quantum number, which we call the $D$-parity.
All known particles are $D$-even, while the $D$ is $D$-odd.
\item
In addition to  $D$,  other $D$-odd particles
exist:  a charged $D^\pm$ and (sometimes) a neutral $D^A$, with
the same spin $s_D$ and with masses
$M_+,\, M_A>M_D$.
The $D$-parity conservation ensures stability of the
lightest $D$-odd particle $D$.

\item These $D$-particles interact with the SM particles
via the Higgs boson $DDh$, $D^+D^-h$, $D^AD^Ah$ and
via the covariant derivative in the kinetic term of the Lagrangian.
These are the gauge interactions $D^+D^-\gamma$, $D^+D^-Z$,
$D^+DW^-$, $D^+D^AW^-$, $D^ADZ$ with the standard electroweak
couplings $g$, $g'$ and $e$    (coupling to  $Z$ can be added by a mixing factor $\mu_M\le 1$, deviation from 1  appears due to possible mixing of $D$ with other $D$-odd  neutrals).
\end{Enumerate}

\bu \  The first example of such model provides well known MSSM (see e.g. \cite{MSSMdark1}-\cite{WILC-2}) for specific set of parameters. Here  our term $D$-parity means $R$-parity. For the considered set of parameters,  $D$ is the lightest neutralino $\chi_1^0$, the  heavier neutralino $\chi_2^0$ can play role $D^A$ and the next in mass $D$-odd particle is the lightest chargino,  spin of these $D$-particles $s_D=1/2$. The other $D$-odd particles (in particular, sleptons and squarks)
are supposed to be heavier  than the ILC beam energy $E$.

\bu \  The second example of such models provides the Inert Doublet Model  (IDM) (see e.g. \cite{inert-1}-\cite{Lundstrom:2008ai-3} and Appendix~\ref{secinert}). That is the $Z_2$ symmetric Two Higgs Doublet Model, containing two scalar doublets $\phi_S$ and $\phi_D$. The "standard" scalar (Higgs) doublet $\phi_S$ is responsible for electroweak symmetry breaking and the masses of fermions and gauge bosons just as in the Standard Model (SM). The second  scalar doublet $\phi_D$ doesn't receive  vacuum expectation value  and doesn't couple to fermions. In this model the $D$-parity conservation is ensured by a $Z_2$ symmetry, four degrees of freedom of  the Higgs doublet $\phi_S$ are the same as in the SM: three Goldstone modes and  the standard Higgs boson $h$.  All the components of the scalar doublet $\phi_D$ are realized as the massive $D$-particles: two charged $D^\pm$ and two neutral ones $D$, $D^{A}$ with masses $M_+$, $M_D$, $M_A$ respectively with $M_+,\,M_A>M_D$. IDM contains no other $D$-odd particles. All $D$-particles have spin $s_D=0$.

A possible value of mass $M_D$ is limited by  stability of $D$
during the Universe existence \cite{Dolle:2009fn-1}-\cite{PDG}.   The
non-observation of the processes $e^+e^-\to D^+D^-$ and $e^+e^-\to
DD^A$ at   LEP gives   $M_+> 90$~GeV and  $M_A>100$~GeV (at
$M_A-M_D>10$~GeV) \cite{Lundstrom:2008ai-1}-\cite{Lundstrom:2008ai-3}.   Limitations for
masses  of neutralino and chargino can be found in
\cite{PDG}. For IDM,  limitations for parameters   of $Z$-peak,
$S$ and $T$ results in  (\cite{Lundstrom:2008ai-1}-\cite{Lundstrom:2008ai-3}, \cite{PDG})
 \be
|\Delta T|=2.16 \left(\fr{M_+-M_D}{v}\right) \cdot
\left|\fr{M_+-M_A}{v}\right|<0.15\,\label{Tlim}
 \ee
 (with $v=246$~GeV -- vacuum expectation value of Higgs field).
Further we
will have in mind  $ M_D\lesssim 80~\mbox{GeV}$ and assume
$M_+-M_D,\, M_A-M_D>20$~GeV.

\subsection{The problem}

The neutral and stable $D$ can  be produced and detected
via  production $D^\pm$ or $D^A$ and subsequent decay
$D^\pm\to DW^\pm$, $D^A\to DZ$ with either
on shell (real) or off shell $W^\pm$ or $Z$. The off shell $W$ emerges
as a  $q\bar{q}$
pair (dijet\fn{We use term
"dijet"  for all products of $W$ decay in $q\bar{q}$ mode -- that are 2 quark jets or 2 quark jets plus gluon jet(s) or few hadrons for off shell $W$ with small effective mass.}) or \ $\ell\nu$, having the same quantum numbers as $W$
but with an effective mass $M^*<M_W$. From now on, $W$ or $Z$ refers to any of these two cases.

 To discover \ the DM particle, one  needs to specify such processes with a clear signature. The $e^+e^-$ Collider ILC/~CLIC  provides an excellent opportunity  for this task    (see, e.g., \cite{TESLA-1}, \cite{TESLA-2}) in the process $\epe\to D^+D^-$ with a clear signature, see Eqs.~\eqref{sign+W} and \eqref{signWZ1} below. The cross section of this process  is a large fraction of  the total cross section of \epe annihilation,  sect.~\ref{secxsecpm}.

The masses $M_+$ and $M_D$ could be found via the edges of the
energy distribution  of dijets, originating from $W$ from decay
$D^\pm\to DW^\pm$, sect.~\ref{secWdistr}, \ref{secleptlightA}
(see \cite{WILC-1}-\cite{WILC-2} for MSSM and \cite{Lundstrom:2008ai-1}-\cite{Lundstrom:2008ai-3} for
IDM). However, this method cannot provide a good accuracy in
measuring the mass. Indeed, the individual jet energy measurement
suffers from a sizable uncertainty.  In particular, this
uncertainty smoothes the lower edge in the dijet energy
spectrum.

On the contrary, the lepton energy can be measured much more
precisely.  In this paper we show, first, that  the energy distribution of leptons has singular points whose positions are kinematically determined. Measuring  positions of these singularities will allow, in principle, to  determine masses  $M_D$ and $M_+$ with good precision (sect.~\ref{secleptlight+}, \ref{secleptlightA}).  In contrast to \cite{WILC-1}-\cite{WILC-2}, \cite{Lundstrom:2008ai-1}-\cite{Lundstrom:2008ai-3}, our \ description is suitable for different models.

Moreover, we present a simple method for measuring spin of DM particles in
these very experiments.

The discussed problem differs strongly from that for the case when the lightest charged D-odd particle is slepton (another set of parameters of MSSM). In the latter case DM particles are produced via  slepton pair,  $e^+e^-\to \tilde{\ell}^+\tilde{\ell}^-\to \ell^+\ell^-\chi_0\chi_0$.  First of all, signature of this process is quite different from that one in our problem \eqref{sign+W}, \eqref{signWZ1}. Second,  the energy of observable lepton -- decay product  of slepton is  measurable well {\it in each individual event}, in difference with our case, when similar product of decay, $W$,  is seen as dijet or lepton plus neutrino with badly measurable energy in each individual event. Therefore, the approach used in the analysis of slepton production  (cf. \cite{selectron-1}-\cite{selectron-3}) cannot be applied directly to our problem.

The overall picture is summarized in sect.~\ref{secsum}. Short conclusion is given in sect.~\ref{secconcl}.

In the Appendix B we discuss the potential of the process $\epe\to
DD^A\to DDZ$ for similar problems, for completeness.  In
contrast with previous studies, we find that this
potential is not too high.

In the Appendix C we consider possible background processes and
show that the most of them  can be neglected at the analysis.

\subsection{Scale of cross sections}

We express discussed cross sections via
\be
\sigma_0\equiv \sigma(e^+e^-\to \gamma\to\mu^+\mu^-) =4\pi\alpha^2/3s\,.\label{sig0}
\ee
The total cross section of $e^+e^-$ annihilation at ILC for \
$\sqrt{s}\equiv 2E>200$~GeV is  $\sim 10\;\sigma_0$.
 The  annual integrated
luminosity $\cal L$ for  the ILC
project \cite{TESLA-2} gives
\be
{\cal L}\sigma_0\sim 3\cdot 10^5.\label{annlum}
\ee

 The process $e^+e^-\to D^+D^-$ represents a significant
fraction of all $e^+e^-$ annihilation events -- see \eqref{crsec},
\eqref{compcrsec}, Fig.~\ref{xsec+} and Table~\ref{tabA}. With the
luminosity \eqref{annlum}, the annual number of events of
discussed type will be $(0.6\div 3)\cdot 10^5$, depending on
$M_+/E$ and  $s_D$, and about $1/3$ of them (in the mode
with $e$ or $\mu$ plus dijet)   are suitable for
our analysis.

\section{The process \ \ $\pmb{e^+e^-\to D^+D^-}$}\label{secmain}

Note before all that the energies, $\gamma$-factors and velocities of
$D^\pm$  are
 \be
 E_\pm=E=\sqrt{s}/2,\quad \gamma_\pm=E/M_+, \quad \beta_\pm= \sqrt{1-M_+^2/E^2}. \label{cmkinpm}
 \ee

\subsection{The signature}

\bu {\bf    If \ \ $\pmb{M_A>M_+}$ or $\pmb{D^A}$ is absent}, once
produced, particles $D^\pm$ decay fast (with a unit probability)
to $DW^\pm$, \be \epe\to D^+D^-\to DDW^+W^-\,.\label{DDWWmain} \ee
The observable states are decay products of $W$  with a large
missing transverse energy \ $\missET$ carried away by the
invisible $D$-particle, and the missing mass of particles escaping
observation $M(\missET)$ is large. In contrast to the LHC, where a
large flux of low $p_\bot$ particles demands an additional
$p_\bot$ cut off, at $\epe$ LC such particles are absent.

Therefore, the  signatures of the process  in the modes
suitable for observation are
  \bes\label{sign+W}
   \be
\boxed{\mbox{\begin{minipage}{0.73\textwidth} \cl{$\epe\to DD(W\to q\bar{q})(W\to q\bar{q})$: \ \ \ \ \ \ \ \  Two dijets
+ {\large\it nothing},} with energy of each dijet  \
  $<E$,  with large   \  $\missET$  and large
$M(\missET)$.   \end{minipage} }}
\label{sign+WB}
\ee
  \be
\boxed{\mbox{\begin{minipage}{0.83\textwidth} \cl{$\epe\to DD(W\to \ell\nu)(W\to q\bar{q})$: \ \ \ \ \ \   One dijet $+$ $ e$ or $\mu$
+ {\large\it nothing},} with energy of each dijet or lepton  \
  $<E$,  with large   \  $\missET$  and large
$M(\missET)$.   \end{minipage} }}
\label{sign+WA}
\ee
 \ees

At $M^*>5$~GeV, the branching ratios  for different
channels of $W$ decay are roughly identical for on-shell $W$
\cite{PDG} and off-shell $W$.  In particular, the fraction
of events with signature \eqref{sign+WB} \  is
$0.676^2\approx 0.45$. The fraction of events with signature \eqref{sign+WA} is $2\cdot 0.676\cdot2\cdot (1+0.17)\cdot
0.108\approx 0.33$ (here 0.17 is a fraction of $\mu$ or $e$ from
the decay of $\tau$). At  $M^*<5$~GeV,
$BR(e\nu)$ and $BR(\mu\nu) $ increase, while the dijet becomes
a set of a few hadrons.

\bu {\bf If $\pmb{M_+>M_A}$}, when analysing the main process
$\epe\to D^+D^-$, {\it one more decay channel is added, $D^\pm\to D^A W^\pm\to DZW^\pm $}. Its branching ratio $B= BR(D^+\to D^AW^+)$\lb is typically less than 0.5 (see discussion in sect.~\ref{secleptlightA}). Particle $D^A$ decays fast to $DZ$,
creating  new cascades $e^+e^-\to D^+D^-\to DW^+D^AW^-\to DD W^+W^-Z$,\lb  $e^+e^-\to D^+D^-\to D^AW^+D^AW^-\to DD W^+W^-ZZ$.  As a result,  the signature of the processes $e^+e^-\to D^+D^-$  in the modes suitable for observation contains both \eqref{sign+W}  and processes with decay $W$'s or $Z$'s in the mentioned cascades:
 \be
\boxed{\mbox{\begin{minipage}{0.86\textwidth}
 \cl{  $4\div 1$ dijets and $0\div 5$ leptons
 with large    $\missET$  and large
$M(\missET)$  + {\large\it nothing}.}\end{minipage}
}}\label{signWZ1}
 \ee
{Note that the processes with invisible decay $Z\to \nu\bar{\nu}$  (we denote these states as $Z_n$, their $BR=20\%$ ) have  signature \eqref{sign+W}.}

\subsection{$\pmb W$   energy distribution in \ \ $\pmb{\epe\to D^+D^-\to DDW^+W^-}$}\label{secWdistr}

Here we consider the energy distribution of  $W$  with an effective mass $M^*$.
At each value of $M^*$, we have in the rest frame of $D^\pm$ a
two-particle decay $D^\pm\to DW^\pm$ with\fn{We denote quantities in the rest system of $D^\pm$ and in the Lab system by using superscripts $r$ and $L$  respectively, additional superscript $+$ or $-$ corresponds to upper or lower value of this quantity. Subscripts $on$ or $off$ correspond to on shell or off shell $W$'s, subscripts $p$ or $k$ mark values, correspondent to peak or kink. Other subscripts and superscripts look evident.}
 \bear{c}
\!\!E_{W}^r(M^*)\!=\!\fr{M_+^2 +M^{*2}- M_{D}^2}{2M_+},\;\; p^r_{W}(M^*)\!=\!\fr{\Delta(M_+^2,M^{*2},M_{D}^2)}{2M_+},\\[2mm] \Delta(s_1,s_2,s_3)^2=s_1^2+s_2^2+s_3^2-2s_1s_2-2s_1s_3-2s_2s_3.
\eear{rkinW}

Denoting  by $\theta$ the $W^+$ escape angle in the $D^+$ rest frame
with respect to the direction of $D^+$ motion in the laboratory frame
and using $c\equiv\cos\theta$, we find the energy of \ $W^+$ \ in
the laboratory frame  as $E_W^L=\gamma_\pm(E_{W}^r+c\beta_\pm p_{W}^r) $.
Therefore, at given $M^*$, the energy  $E^L_W$  of  $W$   lies within the
interval $\gamma_\pm(E_{W}^r\pm \beta_\pm p_{W}^r)$.

In particular, {\bf at
$\pmb{M_+-M_D>M_W}$} we have $M^*=M_W$, and  the  kinematical edges of the
$W$ energy distribution are
 \be
E^{L,\pm}_{W,on}\!=\!\gamma_\pm(E_W^r(M_W)\!\pm\!\beta_\pm
p^r_W(M_W)).\label{EPW}
 \ee

{\bf At  $\pmb{M_+-M_D<M_W}$} we have $0\le M^*\le M_+-M_D$,
and obtain similar edges, which are different  for each value of
$M^*$. The absolute upper and lower bounds on the energy distribution of
$W$ are attained at $M^* = 0$, they are equal to
 \be
E^{L,\pm}_{W,off}=
E\,(1\pm \beta_\pm)\left(1-M_D^2/M_+^2\right)/2
\;.\label{ELWoff}
\ee
At the  highest value $M^*=M_+-M_D$ we have
$p_W^r=0$, and an interval \eqref{EPW} is reduced to a point,
where the entire $W$ energy distribution has a maximum (peak) of
\be
E^L_{W,p}\equiv
E^{L, \pm}_W|_{(M^*=M_+-M_D)}=E\left(1-M_D/M_+\right).\label{peakW}
 \ee

\subsection{
Single lepton  energy distribution in\\
$\pmb{\epe\to D^+D^-\to DD W^+W^-\to DD q\bar{q}\,\ell\nu} (\ref{sign+WA})$
}\label{secleptlight+}

The fraction
of such events  for each separate lepton, $e^+$, $e^-$, $\mu^+$ or
$\mu^-$,
is about 0.08,  their sum is about 0.33 of the total cross section
of the process.
We will speak, for\ \  definiteness, $\ell=\mu^-$ and neglect the muon mass.

Note that in the laboratory frame, for a $W$ with some energy $E_W^L$,
its $\gamma$-factor and the
velocity  are $\gamma_{WL}=E_W^L/M^*$ and
$\beta_{WL}\equiv \sqrt{1-\gamma_{WL}^{-2}}$.

We study the  distribution\fn{We find useful, to mark in the argument of this distribution also masses of produced
$D$-particle $M_+$ and $D$-particle appeared in its decay $M_D$.} of muons over its energy $\vep$,
$d\sigma^\mu(\vep|M_+,\,M_D)/d\vep $.
We show that this distribution  has singular points, whose positions
are kinematically determined, i.e. model independent.

a) If ${M_+-M_D>M_W}$ we have $M^*=M_W$, and  the muon energy and
momentum
in the rest frame of $W$ are $M_W/2$.  Just as above,
denoting by $\theta_1$ the escape angle of $\mu$ relative to the
direction of the $W$ in the laboratory frame and using
$c_1=\cos\theta_1$,
we find that the muon energy in the laboratory frame is
$\vep=\gamma_{WL}\left(1+c_1\beta_{WL}\right)(M_W/2)$. Therefore,
for these muons $\vep^+(E^L_W)\geqslant \vep \geqslant \vep^-(E^L_W)$
where\\ \cl{$ \vep^+(E^L_W)= E_W^L(1+\beta_{WL})/2=(E_W^L+\sqrt{(E_W^L)^2-M_W^2})/2$} and $\vep^-(E^L_W)=M_W^2/\left(4\vep^+(E^L_W)\right)$.

It is easy to check that the interval corresponding to energy
$E_{1W}^L<E_W^L$ is located entirely within the interval,
correspondent to  energy $E_W^L$. Therefore, all muon energies lie
within the interval determined by the highest value of $W$ energy:
 \bear{c}
 \vep^+ \geqslant \vep\geqslant \vep^-\equiv \fr{M_W^2}{4\vep^+},\;\;\mbox{where}
 \;\;  \vep^+\equiv
\vep^+(E^{L, +}_{W,on})=
\fr{
E^{L, +}_{W,on}+\sqrt{(E^{L, +}_{W,on})^2-M_W^2}}{2}\,.
\eear{Emularge}
(Note that $E^{L,+}_{W,on}=\vep^++M_W^2/(4\vep^+)$.)

With a shift of  $\vep$ from these boundaries inwards, the density of
states in the $\vep$ distribution grows  monotonically due to
contributions
of smaller values  $E^L_W$ up to values $\vep^\pm_k$, corresponding
to the lowest value of $W$ energy $E^{L, -}_{W,on}$:
 \bear{c}
  \vep^-_k\equiv
\vep^-(E^{L, -}_{W,on})=
\fr{
E^{L, -}_{W,on}- \sqrt{(E^{L, -}_{W,on})^2-M_W^2}}{2},\;\;\;
\vep^+_k\equiv
\vep^+(E^{L, -}_{W,on})=\fr{M_W^2}{4\vep^-_k}
.\eear{Emuin}
In these  points  the  energy distributions of muons has kinks.
Between these kinks, the $\vep$-distribution is approximately flat.

\begin{figure}
\includegraphics[height=0.2\textheight,width=0.48\textwidth]{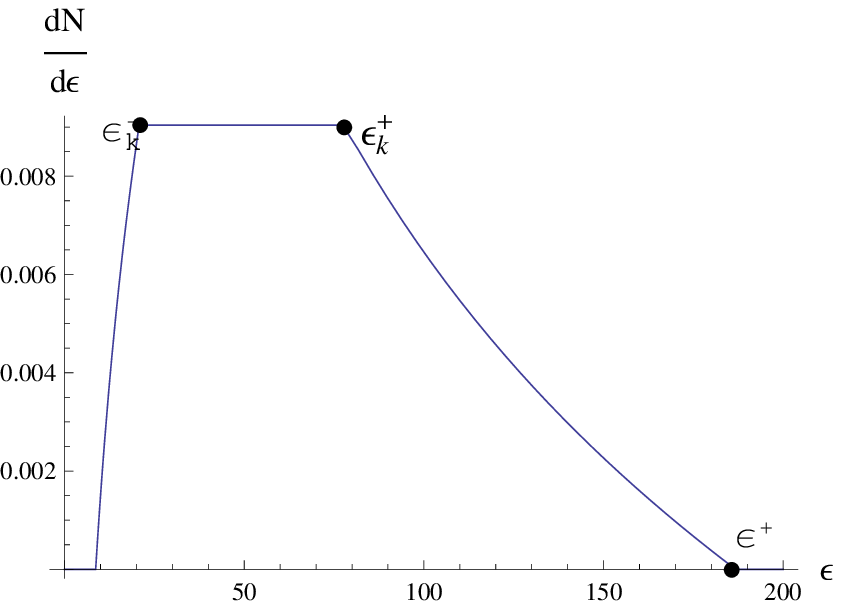}
\hspace{7mm}\includegraphics[height=0.2\textheight,width=0.48\textwidth]{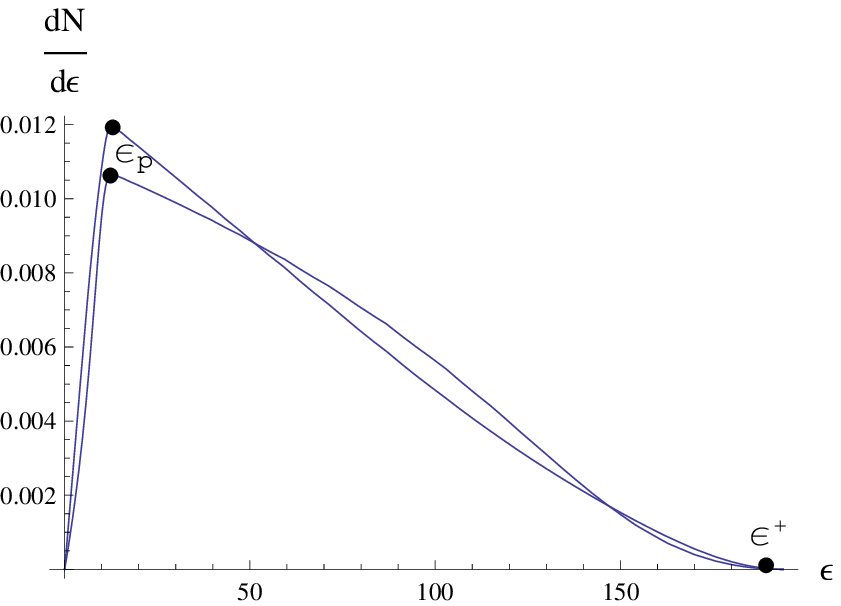}
\caption{\it Distributions $dN/d\epsilon\equiv
(1/\sigma)d\sigma/d\epsilon$    at $E=250$~GeV, $M_D=50$~GeV for $M_+=150$~GeV -- the case with $M_+-M_D>M_W$ (the right plot)
and for $M_+=120$~GeV -- the case with $M_+-M_D<M_W$ (the left plot). In the latter case, the higher and lower peaks are for $s_D=0$
and $s_D=1/2$, respectively.} \label{singlmufig}
\end{figure}

Figure~\ref{singlmufig}, the left plot, shows
the energy distribution of muons for the case of the matrix element independent
of $\theta_1$. Since positions of kinks are kinematically determined, it is not surprising that
calculations for distinct models (containing different angular dependence)
demonstrate variations in shapes but do not perturb the position of kinks.

b) If ${M_+-M_D<M_W}$, the $D^\pm$ decays to $DW^{*\pm}$, where
$W^{*\pm}$
is an off-shell $W$ with an effective mass $M^*\leqslant
M_+-M_D$. The calculations for each $M^*$ similar to shown above
demonstrate that the muon energies  are within the interval
appearing at $M^*=0$:
 \be
\left(\vep^-=0;\;\vep^+= E^{L,+}_{W,off}
\,\right).
 \label{emustarmax}
 \ee
Similarly to the preceding discussion, the increase of $M^*$
shifts the interval boundaries inwards. Therefore, the muon energy
distribution increases monotonously from the outer bounds up to the
maximum (peak) at
$M^*=M_+-M_D$ (cf. \eqref{peakW}):
\be
\vep_p=E(1+\beta_\pm)\left(1-M_D/M_+\right)/2.
 \label{emustarmin}
 \ee

To get an idea about the shape of the peak, we use the
distribution of $W^*$'s (dijets or $\ell\nu$ pairs)  over the
effective masses $M^*$ which is given by the spin-dependent factor $R_{s_D}p^*dM^{*2}$:
 \bear{c}
R_0=\fr{p^{*2}}{(M_W^2-M^{*2})^2}\,,\quad
\!\!\!\!R_{1/2}\!=\!\fr{(M_+^2\!+\!M_D^2\!-\!M^{*2})(2M_W^2\!+\!M_+^2\!+\!M_D^2)\!-\!4M_+^2M_D^2}{(M_W^2-M^{*2})^2M_W^2}.
\eear{offshelmass}

The density of muon states in energy  is calculated by
convolution of kinematically determined distribution with
distribution \eqref{offshelmass}.  Neglecting the angular dependence of
the matrix element, we obtain the result in form of
Fig.~\ref{singlmufig}, right plot. One can see that the discussed peak is
sharp enough for both values of spin $s_D=0$ and $1/2$.

Characteristic values for singular points in  the energy
distributions of muons (kink and peak) together with similar points
for the energy distributions of $W$ (dijets) are given in Table~\ref{tab+} .
\begin{table}[htb]
\caption{\it The singular point energies of lepton and  dijet
in $\epe\to D^+D^-\to DDq\bar{q}\ell\nu$  (in GeV) at $M_D=50$~GeV.}\label{tab+}\vspace{2mm}\centering
\begin{tabular}{|c|c||c|c|c|c | c|c|}\hline
$E$ &$M_+$&$\vep^+$&$\vep^-_k $&$\vep^+_k $&$\vep_p$&$E^L_{Wp} $&$E^{L,+}_W$\\ \hline
250&150&186.3&20.8&77.8&-&-&195.4\\ \hline
250&200&184.9&34.9&46.3&-&-&193.6\\ \hline
250&80&148.3&-&-&91.3&93.8&148.3\\\hline
100&80&78&-&-&30&37.5&78\\\hline
\end{tabular}
\end{table}


{\bf The cascade $\pmb{D^-\to DW^-\to D\tau^-\nu\to
D\mu^-\nu\nu\nu}$} modifies the spectra just discussed. The energy
distribution of $\tau$ produced in the decay $W\to \tau\nu$ is the
same as that for $\mu$ or $e$, discussed above (within the accuracy of $\sim M_\tau/M^*$). Once produced, $\tau$ decays to $\mu\nu\nu$ in 17 \% of  cases
(the same for decay to $e\nu\nu$). These muons are added to those discussed above.

In the $\tau$ rest frame, the energy of muon is $E_\mu^\tau=y\,M_\tau/2$ with $y\leqslant 1$. The energy spectrum of muons  is $dN/dy=2(3-2y)y^2$ (see textbooks). This spectrum and the distributions obtained above are converted into the energy distribution of muons in the Lab frame. It is clear that
this contribution  is strongly shifted towards the soft end of
the entire muon energy spectrum.

The resulting distribution retains the upper boundary of the energy distribution of muons $\vep^+$ \eqref{Emularge}, \eqref{emustarmax}.
Numerical examples  show that here the
upper kink is smeared, while lower kink $\vep_k^-$ become even more sharp without shift from position \eqref{Emuin} in wide region of masses $M_+$ and $M_D$. The position of peak \eqref{emustarmin} is also shifted weakly.

\subsection{Additional decay channels at  $\pmb{M_+>M_A}$
}\label{secleptlightA}

At $M_+>M_A$ the decay $D^\pm\to D^AW^\pm \to DZW^\pm$ become possible and   the processes $\epe\to D^+D^-\to W^+W^-D^AD\to DDW^+W^-Z$, etc. with signature \eqref{signWZ1} should be taken into account.

The total probability of $D^+$ decay to $DW^+$ and  $D^AW^+$
equals 1. The decay\lb $D^\pm \to D^AW^\pm$ is described by the same
equation as $D^\pm \to DW^\pm$, but with other kinematical
factors since $M_A\neq M_D$. In the IDM the probability of this new decay is
lower than that without $D^A$ due to smaller final phase space,
i.e. $B= BR(D^+\to D^AW^+)<0.5$. In the MSSM value of $B$ depends additionally on the mixing angles. We assume that in general case $B\lesssim 0.5$.

Below we limit ourself by the study of processes with signature \eqref{sign+WA}, (\ref{signWZ1}a). Unfortunately, some of new  processes with intermediate $D^A$ look as those with signature \eqref{sign+W} since large fraction (20\% ) of decays of $Z$ is  invisible
($\nu\bar{\nu}$ final states). We denote these states of $Z$
as $Z_n$.

Let us consider in more detail production of an observed state with signature (\ref{sign+WA}), (\ref{signWZ1}a) $\mu^-$ {\it + (1-2) dijets  +  nothing}.
This state can be obtained from two  different cascades.

1) The  cascade $D^-\to DW^-\to D\mu^-\nu$. The energy distribution
of $\mu^-$ here reproduces $d\sigma^\mu(\vep|M_+,\,M_D)/d\vep$,
discussed in sect.~\ref{secleptlight+} with an additional factor $(1-B)$.

2) Cascade $D^-\to D^AW^-\to DZ_n\mu^-\nu$. Since couplings $D^-DW^-$
and $D^-D^AW^-$ differ by a phase factor only (and perhaps mixing angle factors), the energy distribution
of $\mu^-$ in this case  is described by the same dependence $d\sigma$
but with the change\lb $M_D\to M_A$, the corresponding contribution to the
entire energy distribution is\lb $0.2B
d\sigma^\mu(\vep|M_+,\,M_A)/d\vep $. For brevity we will write
$d\sigma^\mu(\vep|M_+,\,M_D)\to d\sigma^\mu_W$ and
$d\sigma^\mu(\vep|M_+,\,M_A)\to d\sigma^\mu_{WZn}$. The
resulting energy distribution is
\be
d\sigma_{tot}^\mu/d\vep=(1-B)d\sigma_W^\mu/d\vep
+0.2Bd\sigma_{WZn}^\mu/d\vep\,. \ee

The shape of the distribution $d\sigma_{WZn}^\mu/d\vep$ is similar to
that for $d\sigma_W^\mu/d\vep$,  but with
different positions of kinks and (or) peak. As $M_A>M_D$, these new kinks
and (or) peak are situated below similar points for
$d\sigma_W^\mu/d\vep$. Since this contribution is much smaller
than the main contribution $d\sigma_W^\mu/d\vep$ (with
the overall ratio $0.2B/(1-B)$ at $B<0.5$), it only results in a weak
reshaping of the full energy distribution as compared with
distributions $d\sigma_W^\mu/d\vep$.

Note that in the case $M_A\approx M_D$ the distributions
$d\sigma_{WZn}^\mu/d\vep$ and $d\sigma_W^\mu/d\vep$ are close to
each other, and $d\sigma_{tot}^\mu/d\vep\propto d\sigma_W^\mu/d\vep$.
In the opposite degenerate case $M_+\approx M_A$, the quantity $B\ll1$
and the influence of the intermediate $D^A$ state  on the result is
negligible. (Such very cases are widely discussed in context of MSSM).

\section{The overall picture}\label{secsum}

{\bf Observation} of  events with signature \eqref{sign+W}, \eqref{signWZ1}  will be a clear {\it signal} for DM particle candidates. The non-observation of such events will allow to find lower limits for masses $M_+$, like \cite{Lundstrom:2008ai-1}-\cite{Lundstrom:2008ai-3}. One can hope that these limits will be close to the beam energy $E$.\lb\vspace{-5mm}

At $M_+<E$, the cross section $\epe\to D^+D^-$ is
a large fraction of  the total cross section of \epe annihilation,
and it  makes this observation a very realistic task.

\subsection{Distortion of the obtained results}\label{secdist}

A more detailed analysis reveals two sources of distortion of the obtained results (we neglect them in our preliminary analysis).

1. The  final width of $W$ and $D^\pm$ ($Z$ and $D^A$) leads to a
blurring  singularities derived. This effect increases with
the growth of $M_+-M_D$.

2. The energy spectra under discussion will be smoothed due to   QED
initial state radiation (ISR), final state radiation (FSR)
and beamsstrahlung (BS). The ISR and FSR spectra are machine
independent, while BS spectrum is   specific for each machine (but well known during operations). This smoothing decreases accuracy in measuring of
masses. However, the precise knowledge of mentioned spectra allows to solve   the problem about restoration original accuracy by means methods of
deconvolution in so called "incorrect inverse problem".  This work and
the estimates of the  range  where masses and spins can be determined with reasonable accuracy will be  the subject of the forthcoming paper.

\subsection{Masses}

{\bf Masses $\pmb{M_D}$ and $\pmb{M_+}$.} In a well known approach, one measures edges in the energy distributions of dijets, representing $W$ in the decay $D^\pm\to DW^\pm$  \cite{WILC-1}-\cite{WILC-2}.  However,  the
individual jet energies and consequently, effective masses of dijets
cannot be measured  with a high precision. The observed lower edge of the $W$ energy distribution in the dijet mode and the position of a peak in this distribution \eqref{peakW} are  smeared by this uncertainty. One can only hope for a sufficiently accurate measurement of the upper edge of the $W$ energy distribution, $E^{L,+}_{W}$ \eqref{EPW}, \eqref{ELWoff}.

We suggest to extract the second quantity for description of masses
from the  lepton energy
spectra. The lepton energy is measurable  with a high accuracy.
We found above that the singular points of the energy distribution
of the leptons in the final state   with signature \eqref{sign+WB}
are kinematically determined, and therefore can be used for a mass
measurement.

{\it M1)} If a $D^A$ particle is absent or $M_A>M_+$, the
results \eqref{Emularge}-\eqref{emustarmin} describe the energy
distributions completely. The shape of the energy distribution of
leptons (with one peak or two kinks) allows to determine
which case is realized, $M_+-M_D>M_W$ or\lb $M_+-M_D<M_W$.

At $M_+-M_D>M_W$, the positions of upper edge in the dijet energy distribution\lb $E^{L,+}_{W,on}$  \eqref{EPW} and
the lower kink in the muon energy distribution $\vep^+_k$
\eqref{Emuin} give us two equation necessary for
determination of $M_D$ and $M_+$. We reproduce these equations for clarity
\bear{c}
E^{L,\pm}_{W,on}=\fr{E}{M_+}\left(E_W^r\pm \beta_+ p_W^r\right),\;\;
 \vep^-_k=
\fr{
E^{L, -}_{W,on}- \sqrt{(E^{L, -}_{W,on})^2-M_W^2}}{2},\;\;\mbox{where }\\[3mm]
 E_{W}^r\!=\!\fr{M_+^2 +M_W^2- M_{D}^2}{2M_+},\;\; p^r_{W}\!=\!\fr{\Delta(M_+^2,M_W^2,M_{D}^2)}{2M_+},\;\; \beta_+=\sqrt{1-\fr{M_+^2}{E^2}}.
\eear{massseteqon}

At $M_+-M_D<M_W$,
two  similar equations are provided by the position of
the upper edge in the dijet energy distribution $E^{L,+}_{W,off}$  \eqref{ELWoff} and the peak in muon energy distribution $\vep_p$ \eqref{emustarmin}.

In both cases the position of the upper edge in the dijet
energy distribution $E^{L,+}_{W,on}$ or $E^{L,+}_{W,off}$ should
be extracted from all events with signature \eqref{sign+W},
\eqref{signWZ1},  the position of the lower kink in the muon
energy distribution $\vep^+_k$ or peak $\vep_p$ can be extracted
from  events with signature \eqref{sign+WA} only.

{\it M2)} The signal of realization of  the
inequality ${\pmb M_A<M_+}$ will be observation of the
process $\epe\to DD^A$, having signature \eqref{DDAZ}. In this
case  the position of the upper edge in the dijet energy
distribution is the same as in previous case. The position of
lower edge in the dijet energy distribution is either shifted or
smeared, in this case  the method of \cite{WILC-1}-\cite{WILC-2}
becomes completely inapplicable.   The entire energy
distribution of muons in the observed state {\it $\mu$ +1  or 2
dijet + nothing} was described in the sect.~\ref{secleptlightA}.
It was shown there that taking into account a new decay channel
$D^-\to D^AW^-\to DZ_n\mu^-\nu $ changes  the position  of  the
main singularities in the muon energy spectrum very weakly.
Therefore the above mentioned procedure for finding $M_+$ and
$M_D$ can be used in this case as well.

The opportunity to extract new singularities from the data,
related to $d\sigma_{WZn}^\mu/d\vep$ (and giving additionally
$M_A$), requires a separate study (see also analysis in
Appendix B).

\subsection{Spin of $\pmb D$-particles $\pmb{s_D}$}\label{secxsecpm}

\begin{figure}[htb]\centering
\includegraphics[width=0.75\textwidth,height=0.25\textheight]{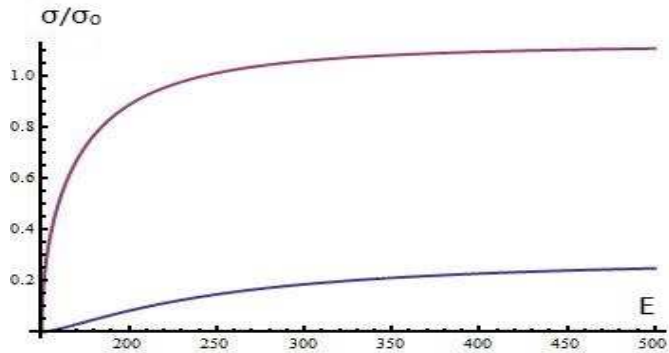}\vspace{-4mm}
\caption{\it
The upper    curve -- \ for $s_D=1/2$, the lower -- for $s_D=0$; $M_+=150$~GeV.}\vspace{-7mm}
\label{xsec+}
\end{figure}

\bu  The  amplitude of  the process $\epe\to D^+D^-$
is the sum of model-independent QED diagram (the photon
annihilation), the $Z$ annihilation diagram and -- in some models
-- $t$-channel exchange by other $D$-odd particles. We start with
the description of cross section in the minimal
approximation, taking into account only photon and $Z$
annihilation diagrams. Neglecting terms \ \
$\propto\!\!(1/4-\sin^2\theta_W)$ (described $\gamma -Z$
interference) we have:
\bear{c} \!\!\sigma_{min}(s_D)=\!
\sigma_0\left\{\!\!\begin{array}{lr}
\!\!\beta_\pm\!\left[1+\fr{2M_+^2}{s}
+r_Z\beta_\pm^2\right]&\left(s_D=\fr{1}{2}\right),\\[3mm]
\!\!\beta_\pm^3\!\left[\fr{1}{4}+r_Z\cos^2(2\theta_W)\! \right]&\left(s_D=0\right)
\!,\end{array}\right.
\eear{crsec}
where
$r_Z=\fr{\mu_M}{\left(2\sin(2\theta_W)\right)^4(1-M_Z^2/s)^2}=\fr{0.124\mu_M}{(1-M_Z^2/s)^2}$, factor $\mu_M\le 1$ is expressed via parameters of possible mixing, etc.
Fig.~\ref{xsec+} and Table~\ref{tabA}
represent dependence of
 $\sigma_{min}(e^+e^-\to D^+D^-)/\sigma_0$ \eqref{crsec} on beam energy $E=\sqrt{s}/2$ for $\mu_M=1$.

\begin{table}[h]\centering
\caption{\it Some values of $\sigma_{min}(e^+e^-\to D^+D^-)/\sigma_0$}
\label{tabA}
\begin{tabular}{|c||c|c|c|c|}\hline
$E$, GeV&100&250&250&250\\\hline
$M_+$, GeV & 80&80&150&200\\\hline
${s_D=0}:$ $\sigma/\sigma_0$&
0.066&0.245&0.162&0.062\\\hline
${s_D=1/2}:$ $\sigma/\sigma_0$&
0.84&1.107&1.02&0.82\\\hline
\end{tabular}
\end{table}

The  cross section of the process is reduced by
contribution of the diagram with $t$-channel exchange by
other $D$-odd particle $D_F$. This decrease is not so strong if
mass of $D_F$ is high enough. For example, if mass of selectron is
more than 250~GeV (condition 2 in sect.~\ref{secintro} and
\cite{PDG}), the cross section for $s_D=1/2$ is reduced  by
a factor $\ge 0.6$,
$\sigma(s_D=1/2)\ge 0.6\sigma_{min}(s_D=1/2)$. Combining with
numbers from Fig.~\ref{xsec+} and Table~\ref{tabA} we obtain
 (for identical masses $M_+$ at  a given beam
energy $E$):
 \be
 \sigma(s_D=1/2)>2\sigma_{min}(s_D=0).\label{compcrsec}
\ee

\bu \ The  experimental value of the $e^+e^-\to D^+D^-$ cross section is
obtained by summing over all processes with signature \eqref{sign+W},
\eqref{signWZ1} (that is about 3/4 of the total cross section). By taking into
account the known BR's for $W$ decay the accuracy of this restoration of $\sigma(e^+e^-\to D^+D^-)$ can be improved.

When masses $M_+$ become known, the cross section
$\sigma_{min}(e^+e^-\to D^+D^-)$ is calculated with reasonable
precision with eq.~\eqref{crsec}.  The strong inequality
\eqref{compcrsec}   allows  to determine spin $s_D$ from
the obtained values of cross sections even with a handful of
well-reconstructed events.

\section{Conclusions}\label{secconcl}

We consider models in which stability of dark matter particles $D$ is ensured by conservation of new quantum number referred to as $D$-parity. Besides these models contain charged particles $D^\pm$ with the same $D$-parity. (Examples -- Inert Doublet Model with scalar $D$-particles and MSSM with $D$-particle of spin 1/2 and $D$-parity equal $R$-parity). In these models we have studied the energy distribution of single lepton in the process like $e^+e^-\to D^+D^-\to DD W^\pm(\to qq)W^\mp(\to \ell \nu)$, having high enough cross section. Simple analysis allows us to establish that this distribution has singular points, kinks, peaks and end points,  which are driven by kinematics only, and therefore are model-independent. Based on this analysis, we propose to use the mentioned distribution at future linear $e^+e^-$ collider ILC, CLIC, etc. for precise measuring of masses of dark matter particles and charged particles $D^\pm$.

This  method  is in several aspects superior to the standard approaches discussed elsewhere.

1) It uses leptons which are copious and can be accurately measured in contrast with jets which individual energy can be measured only with lower precision.

2) These singularities are robust and survive even when superimposed on top of any smooth background.

In addition, even a rough measurement of cross sections with a very clean signature allows to determine spin of DM particles based on the results of mentioned kinematical measurements.

\section*{Acknowledgments}

This work was supported in part by grants RFBR and
NSh-3802.2012.2, Program of Dept. of Phys. Sc. RAS and SB RAS
"Studies of Higgs boson and exotic particles at LHC" and  Polish Ministry of Science and Higher Education Grant N202 230337. I am thankful to A. Bondar, E. Boos, A.~Gladyshev,  A. Grozin,  S. Eidelman, I.~Ivanov, D.~Ivanov, D.~Kazakov, J.~Kalinowski, K.~Kanishev, P.~Krachkov and V.~Serbo for discussions.

\appendix

\section{Inert doublet model (IDM)}\label{secinert}

The IDM the $Z_2$ symmetric Two Higgs Doublet Model, containing two scalar doublets $\phi_S$ and $\phi_D$, in which fermions are coupled to only field $\phi_S$. The $Z_2$ symmetry forbids $\phi_S,\,\phi_D$ mixing. It  is described by the Lagrangian
\begin{equation}
{ \cal L}={ \cal L}^{SM}_{ gf }  + {\cal L}_Y(\psi_f,\phi_S) +\left({\cal D}_\mu\phi_S^\dagger{\cal D}_\mu\phi_S+{\cal D}_\mu\phi_D^\dagger{\cal D}_\mu\phi_D\right)/2-V\, .
\label{lagrbas}
\end{equation}
Here, ${\cal L}^{SM}_{gf}$ describes the $SU(2)\times U(1)$ Standard Model interaction of gauge bosons and fermions.  The ${\cal L}_Y$ describes the Yukawa interaction of fermions $\psi_f$ with only one scalar doublet $\phi_S$, having the same form as in the SM, ${\cal D}_\mu$ is standard covariant derivative,
\bear{c}
V=-\fr{1}{2}\left[m_{11}^2(\phi_S^\dagger\phi_S)\!+\! m_{22}^2(\phi_D^\dagger\phi_D)\right]
\!\!\!+\fr{\lambda_1}{2}(\phi_S^\dagger\phi_S)^2\!+\!\fr{\lambda_2}{2}(\phi_D^\dagger\phi_D)^2+\\[2mm]
+\lambda_3(\phi_S^\dagger\phi_S)(\phi_D^\dagger\phi_D)\!
\!+\!\lambda_4(\phi_S^\dagger\phi_D)(\phi_D^\dagger\phi_S) +\fr{\lambda_5}{2}\left[(\phi_S^\dagger\phi_D)^2\!
+\!(\phi_D^\dagger\phi_S)^2\right].
\eear{baspot1}
All parameters of this Higgs potential are real
and condition of its stability has form
\be
 \lambda_1>0\,, \quad \lambda_2>0,\quad R = \fr{\lambda_3+\lambda_4+\lambda_5}{\sqrt{\lambda_1 \lambda_2}}>-1\,.
\ee
The IDM is realized at
\be
\ls5<0,\quad \ls4+\ls5<0,\quad m_{11}^2>0,\quad \left\{ \begin{array}{cc}
\fr{m_{11}^2}{\sqrt{\lambda_1}}>\fr{m_{22}^2}{\sqrt{\lambda_2}} & at\;\;R>1\\
R\;\fr{m_{11}^2}{\sqrt{\lambda_1}}>\fr{m_{22}^2}{\sqrt{\lambda_2}} & at\;\;|R|<1.\end{array}\right.\label{IDMpot}
\ee
In this case $\la\phi_S\ra=\begin{pmatrix}0\\v\end{pmatrix}$, $\la\phi_D\ra=0$. Therefore, the  doublet $\phi_S$ is responsible for electroweak symmetry breaking and the masses of fermions and gauge bosons just as in the Standard Model (SM). The second  scalar doublet $\phi_D$ doesn't receive  vacuum expectation value  and doesn't couple to fermions.

Four degrees of freedom of  the Higgs doublet $\phi_S$ are such as in
the SM: three Goldstone modes become longitudinal components of the EW gauge
bosons, one  component becomes the standard Higgs boson $h$.  All the
components of the scalar doublet $\phi_D$ are realized as the massive
$D$-particles: two charged $D^\pm$ and two neutral ones $D$, $D^{A}$ with masses $M_+$, $M_D$, $M_A$ respectively.
$$
M_A^2=M_D^2-\ls5 v^2,\quad M_+^2=M_D^2-(\ls4+\ls5)v^2/2.
$$
By construction, $D$-particles possess a new conserved multiplicative quantum number (named here as $D$-parity) and therefore the lightest particle among them can be  a candidate for DM particle.  In this model all $D$-particles have spin $s_D=0$. The other features of IDM can be found in ref.~\cite{inert-1}-\cite{inert-4}.

\section{Process \ \ $\pmb{e^+e^-\to Z\to DD^A\to DDZ}$}\label{secDDA}

One more process leading to production of $D$-odd particles at ILC
is also  observable at\lb $M_A+M_D<2E$ (in particular, at $E>M_+>M_A$):
 \be
e^+e^-\to Z\to DD^A\to DDZ.
 \label{DDZproc}
 \ee
  This process has a clear signature in the modes suitable for observation
\bes\label{DDAZ}
\be
\boxed{\mbox{\begin{minipage}{0.85\textwidth}
The \epe \ or $\mu^+\mu^-$ \ pair   with large \
$\missET$  and large $M(\missET)$ + {\large\it nothing}. The
effective mass of this dilepton  is  $\le M_Z$,
its energy is typically less than $E$. \end{minipage} }}
\label{DDAZA}\ee
\be
\boxed{\mbox{\begin{minipage}{0.8\textwidth} A  quark dijet with large \
$\missET$  and large $M(\missET)$ + {\large\it nothing}. The
effective mass of this  dijet is  $\le M_Z$,
its energy is typically less than $E$. \end{minipage} }}
\label{DDAZB}\ee
\ees

At $M_A<M_+$ the BR for channel with signature \eqref{DDAZA} is 0.06, for the channel with signature \eqref{DDAZB} -- 0.7. We skip channel $Z\to \tau^+\tau^-$ with BR=0.03, 20\% of decays of $Z$ are invisible ($Z\to\nu\bar{\nu}$).

At $M_A>M_+$ BR's for processes with signature    \eqref{DDAZ} become less, since new decay channels $D^A\to D^\mp W^\pm\to DW^+ W^-$ are added with  signature
\be
\boxed{\mbox{\begin{minipage}{0.88\textwidth}  $\epe\to DD^A\to DDW^+W^-$: Two  quark dijets or dijet + single lepton or two leptons in one hemisphere with large \
$\missET$  and large $M(\missET)$ + {\large\it nothing}. The
effective mass of this system is  $\le M_Z$,
its energy is typically less than $E$. \end{minipage} }}
\label{DDAWW}\ee

The cross section of the process $\epe\to DD^A$ is model dependent. In the IDM it is determined unambiguously, in MSSM result depends on mixing angles and on the nature of fermions $D$ and $D^A$ (Dirac or Majorana).
In all considered cases at\lb $\sqrt{s}>200$~GeV this cross section is smaller than $0.1\sigma_0$. Since the BR for events with  signature \eqref{DDAZA} is 0.06, at the luminosity \eqref{annlum} annual number of events with  this signature is  smaller than $2\cdot 10^3$. This number looks insufficient for  kinematical analysis with high enough precision, (but limitations for masses can be obtained (cf. \cite{Lundstrom:2008ai-1}-\cite{Lundstrom:2008ai-3}  for LEP)).

Nevertheless  we describe, for completeness, the energy distributions of $Z$
in this process.  The obtained equations are similar to
\eqref{cmkinpm}, \eqref{rkinW}--\eqref{ELWoff} for new kinematics.

The $\gamma$-factor and velocity of
$D^A$  in c.m.s. for $e^+e^-$ are   \bear{c}
\gamma_A=\fr{s+M_A^2-M_D^2}{4E M_A},\quad
\beta_A=\fr{\Delta(s, M_A^2, M_D^2)}{s+M_A^2-M_D^2}.
\eear{Aprodcm}
For production of $Z$  with an effective mass $M^*$ ($M^*=M_Z$ at $M_A-M_D>M_Z$ and\lb $M^*\le M_A-M_D$ at $M_A-M_D<M_Z$)
in the rest frame of $D^A$
\be
E_Z^r\!=\!\fr{M_A^2 +M^{*2}- M_{D}^2}{2M_A},\qquad p^r_Z\!=\!\fr{\Delta(M_A^2,M^{*2},M_{D}^2)}{2M_A}.\label{rkin}
\ee

At $M_A-M_D>M_Z$ the $Z$-boson energy $E^L_Z$ lies within the interval with edges
 \be
E^{L,-}_{Z,on}\!=\!\gamma_A(E_Z^r\!-\!\beta_A
p_Z^r),\;\;E^{L,+}_{Z,on}\!=\!\gamma_A(E_Z^r\!+\!\beta_A
p^r_Z).\label{EPZ}
 \ee

At $M_A-M_D<M_Z$ similar equations are valid for each value of $M^*$.
Absolute upper and lower edges of the energy distribution of $Z$
are reached at $M^*=0$:
\be
E^{L,\pm}_{Z,off}=\gamma_A(1\pm \beta_a)(M_A^2-M_D^2)/(2M_A)\,.
\label{EPZoff}\ee
The peak in the energy distribution of dilepton appears  at $M^*=M_A-M_D$:
\be
E^L_{Z,p}=\gamma_A(M_A-M_D)\,.\label{EPZpeak}
\ee

{\bf Masses $\pmb{M_D}$ and $\pmb{M_A}$.} At first sight,
measurement of kinematical edges of the dilepton spectrum  \eqref{EPZ} (at $M_A-M_D>M_Z$) gives two equations for $M_D$ and $M_A$, allowing for determination of these masses. At $M_A-M_D<M_Z$, the same procedure can be performed separately for each value of the effective mass of dilepton \cite{Gin10}. In the latter case, the absolute edges of the dilepton energy spectrum \eqref{EPZoff} and the position of the peak in this spectrum \eqref{EPZpeak} could be also used for measuring $M_D$ and $M_A$.

In any case, the upper edge in the dijet energy spectrum $E^{L,+}_{Z}$ \eqref{EPZ}, \eqref{EPZoff} (signature \eqref{DDAZ}) gives one equation, necessary to find $M_A$ and $M_D$. In principle, necessary additional information gives position of lower edge in the dilepton energy spectrum $E^{L,-}_Z$.  However, as it was noted above, the anticipated number of events with signature \eqref{DDAZA} looks insufficient for obtaining precise results. Together with good results for $M_D$ and $M_+$, one can hope to find an accurate value of $M_A$.

\section{Backgrounds}

\subsection*{C1. Background to the  process with signature \eqref{sign+W}}

We show here that the cross sections of possible
background processes (with suitable simple cuts) are $\sim 10\div 100$ times less than the cross section of the signal process and therefore they  can be
neglected at analysis. Note that some our estimates can be corrected due to ISR and beamstrahlung.

$ {BW1}$. The process $\epe\to W^+W^-$ gives the same
final state as those with signature \eqref{sign+W}. However,   many of its features are not permitted by this signature. This fact  allow to exclude the BW1 process from
analysis with  a good confidence applying suitable cuts.

Let us discuss e.g.  the observable mode  $\mu^-+jj+\missET$.

(a) For the process BW1 energy of  each dijet $E_{jj}=E$.

Application of cut in the dijet energy $E_{jj}<E^c$ with large enough $E-E^c$ keeps all dijets from the signal process and leaves only small fraction  of cross section of process BW1.

The dijet energy $E_{jj}$ in  BW1 can be less than $E^c$ only if $W^-$ (seen as $\ell^-\nu$) is strongly off shell with effective mass much higher than $M_W$. The probability of such situation is estimated easily, it is $\delta_{W1a}\approx \Gamma_w M_W/(\pi E(E-E^c))$. The examples considered in the Table~\ref{tab+} allow to use cut $E^c=200$~GeV and this cut  leaves only 0.0012 of total  cross section of BW1 process.

(b) For  BW1 the missing mass  $M(\missET)=0$. Application of cut $M(\missET)>M^c$ with suitable $M^c$ keeps all events of the signal process but diminishes contribution of BW1  in the events with signature (\ref{sign+W}A)   even further.

For dijet+dijet mode total energy of these dijets  in the process BW1 differs from $2E$ only due to inaccuracy of measurements. For the examples considered in the Table~\ref{tab+} this energy deficit should be larger than 100 GeV.

$BW2$. The same (in its content) final state as we consider for signal process can be achieved via mechanism without at least one intermediate $D^\pm$ in $s$-channel, e.g.
\lb \cl{$\epe\to (W^-\to \mu^-\bar{\nu})(W^+\to D(D^+\to DW^+\to Dq\bar{q}))$.} To simplify text of discussion, we will write here about the case $M_A>M_+$ only.

The contribution of this mechanism to the total cross section is
at least in $\alpha$ times less then that of the signal process.
Indeed, in the signal process the value of cross section is given
by the second order process $\epe\to D^+D^-$. It includes the
intermediate decay $D^+\to DW^+$ with probability 1, the  corresponding cross section is $\sim \alpha/s$ (an additional $\alpha$ in the
formal diagram is compensated by the small $D^+$ width $\Gamma^+$ in
the denominator of propagator). In the process BW2 we have  third
order process with decays in final stage (if $M_+-M_D<M_W$ that is
even the fourth order process). The neutrino exchange term
enhances this contribution only logarithmically. If necessary, it
can be reduced by variation of longitudinal polarization of
initial electron and additionally by   the cut in
transverse momentum of muon $p_{\mu\bot}>40$~GeV.

The interference of this BW2 mechanism with the signal one is also
very small. In particular, in the signal process final
leptons ($\ell^-\nu$) and  $d$ form system with effective mass
$M_+\pm(\sim\Gamma^+)$, while in the process BW2 this value of effective mass is
only small part of entire phase space of this system, contributed
to the total cross section.

Therefore, the contribution of mechanism BW2 can be neglected with
accuracy better than 1\%.

$BW3$. $\epe\to DD^A\to  DD^+W^-\to DDW^+W^-$. This background is absent if $M_A<M_+$ or $M_A+M_D>\sqrt{s}$.
If $\sigma(\epe\to DD^A)$ is not small at given $\sqrt{s}$, this
fact will be seen via an observation of the process  $\epe\to DDZ$
\eqref{DDAZ}. The cross section $\sigma(BW3)< \sigma(\epe\to
DDZ)$, i.e.  it is much less than $\sigma(\epe\to D^+D^-\to
DDW^+W^-)$  (roughly, by one order of  magnitude). In
this process all recorded particles move in one hemisphere in
contrast to the with signal process, where they move in
two opposite hemispheres. Therefore, the  contribution of this
background process may be reduced additionally by
application of suitable cuts.

$BW4$. In the   SM processes with an observed state,
satisfying criterion \eqref{sign+W}, large $\missET$ is carried
away by additional  neutrinos. The corresponding cross section is
at least one electroweak  coupling constant squared $g^2/4\pi$ or
$g^{\prime 2}/4\pi$ smaller than $\sigma_0$, with $g^2/4\pi\sim
g^{\prime 2}/4\pi\sim \alpha$. Therefore,  $\sigma(BW3)\lesssim  0.01\sigma(\epe\to D^+D^-)$.

\subsection*{C2. Background to $\pmb{e^+e^-\to DD^A\to DD \ell^+\ell^-}$}

We consider these background processes only for
completeness, since anticipated number of events for the process
itself is not so large. We subdivide these background processes
into 3 groups.

{BZ1.}  $\epe \to ZZ_n$.   At first sight, this  process can
mimic the process $\epe\to DDZ$. However, the dilepton or dijet
in this process  have the same energy $E$ as the
colliding electrons. The criterion \eqref{DDAZ} excludes
such events from the analysis.

The cross section  $\sigma(\epe\to ZZ_n)\sim 0.2\cdot 3r_Z\sigma_0
\ln (s/M_Z^2)$. The variants of this process with off-shell $Z$,
giving another effective mass of observed dijet or dilepton but
with energy close to $E$, has cross section which is
smaller by a factor $\sim \alpha$.

{ BZ2.} Processes with independent production of separate $\ell_i$:\\
(BZ2.1) $\epe\to DDZ\to DD\tau^+\tau^-\to DD \,
\ell_1^+\ell_2^-+\nu's$, \\
(BZ2.2) $\epe\to DD^A\to DDW^+W^-\to
DD\ell_1\bar{\ell}_2\nu\bar{\nu}$,\\
(BZ2.3) $\epe\to D^+D^-\to DDWW\to
DD\ell_1\bar{\ell}_2\nu\bar{\nu}$,\\
(BZ2.4) $\epe\to W^+W^-\to \ell_1\bar{\ell}_2\nu\bar{\nu}$.

In these processes $\epe$, $\mu^+\mu^-$, $e^-\mu^+$ and
$e^+\mu^-$ pairs are produced with identical probability and
identical distributions. Hence,
{\it the subtraction from the $\epe$ and $\mu^+\mu^-$ data
the measured distributions of \ $e^-\mu^+$ and \
$e^+\mu^-$} eliminates a contribution of these
processes from the energy distributions under interest. This
procedure  does not implement substantial
inaccuracies since cross sections of these processes after
suitable cuts  are small enough.

The cross sections of processes (BZ2.1), (BZ2.2) are  small in
comparison with that for $e^+e^-\to \to DDZ\to DD\mu^+\mu^-$. In the process
(BZ2.3) leptons are flying in the opposite hemisphere, in contrast
to the process under study $\epe\to  DDZ\to DD\mu^+\mu^-$, where
the leptons are flying in the same hemisphere. The  cross section
of the process (BZ2.4) is basically large. The application
of cuts $E_{\ell\bar{\ell}}<E$, $M_{\ell\bar{\ell}}\le M_Z$
leaves  less than $(M_Z^2/s)^2\ln(s/M_Z^2)$ part of the cross section. The obtained quantity becomes smaller than that for the signal.

{BZ3. In the  SM processes with the observed state
\eqref{DDAZ}}, the large $\missET$ is carried away by
additional  neutrino(s). The magnitude of the corresponding cross
sections is at least by one electroweak
coupling constant squared $g^2/4\pi$ or $g^{\prime 2}/4\pi$ less
than $\sigma_0$, with $g^2/4\pi\sim g^{\prime 2}/4\pi\sim \alpha$.
Therefore, the cross sections of these processes are at least one
order of magnitude smaller than the cross section
for this signal process.

\end{document}